\documentclass[twoside]{amsart}
\usepackage{latexsym}
\usepackage{amssymb,amsmath,amsopn}
\usepackage[dvips]{graphicx}   %To insert ps figures
\usepackage{color,epsfig}      %To insert ps figures

               {\begin{list}{}{\leftmargin#1\rightmargin#2}\item{}}%
               {\end{list}}
 
\newtheorem{thm}{Theorem}

\newtheorem{prop}{Proposition}
\theoremstyle{definition}
\newtheorem{defn}{Definition}
\newtheorem{algo}{Algorithm}
\newtheorem{rem}{Remark}
\newtheorem{obs}{Observation}

\def\bea{\begin{eqnarray}}
\def\eea{\end{eqnarray}}

\usepackage[normalem]{ulem}

\begin{document}
%\title[Soft cells and foam cells]{Soft cells and the geometry of foams}
\title[Soft cells, Kelvin's foam and the minimal surfaces of Schwarz]{Soft cells, Kelvin's foam and the minimal surfaces of Schwarz}
\author[G. Domokos, A. Goriely,  \'A. G. Horv\'ath \and K. Reg\H os]{G. Domokos, A. Goriely,  \'A. G. Horv\'ath \and K. Reg\H os}

\address{G\'abor Domokos, HUN-REN-BME Morphodynamics Research Group and Dept. of Morphology and Geometric Modeling, Budapest University of Technology,
M\H uegyetem rakpart 1-3., Budapest, Hungary, 1111}
\email{domokos@iit.bme.hu}
\address{Alain Goriely, Mathematical Institute, University of Oxford}
\email{goriely@maths.ox.ac.uk}
\address{\'Akos G. Horv\'ath, HUN-REN-BME Morphodynamics Research Group and Department of Algebra and Geometry, Budapest University of Technology and Economics, H-1111 Budapest, M\H uegyetem rkp 3.}
\email{ghorvath@math.bme.hu}
\address{Krisztina Reg\H os, HUN-REN-BME Morphodynamics Research Group and Dept. of Morphology and Geometric Modeling, Budapest University of Technology,
M\H uegyetem rakpart 1-3., Budapest, Hungary, 1111}
\email{regoskriszti@gmail.com}

%\subjclass{53A05, 53Z05}
\keywords{Tessellation, soft cell, Kelvin cell, Dirichlet-Voronoi cell}
\begin{abstract}

 Recently, we  introduced a new class of shapes, called \textit{soft cells} which fill space as  \emph{soft tilings} without gaps and overlaps while minimizing the number of sharp corners. We introduced the \textit{edge bending algorithm} that deforms a polyhedral tiling into a soft tiling. Using this algorithm,  we proved that  an infinite class of polyhedral tilings can be smoothly deformed into \textit{standard} soft tilings. Here, we demonstrate that certain triply periodic minimal surfaces naturally give rise to non-standard soft tilings. By extending the edge-bending algorithm, we further establish that the soft tilings derived from the Schwarz P and Schwarz D surfaces can be continuously transformed into one another through a one-parameter family of intermediate non-standard soft tilings. Notably, by carrying its combinatorial structure,  both resulting tilings belong to the \textit{first order equivalence class} of the (e2) tiling, i.e. the Dirichlet-Voronoi tiling on the body-centered cubic ($bcc$) lattice, highlighting a deep geometric connection underlying these minimal surface configurations. By requiring identical end-tangents for edges in a first order class, we also define \textit{second order equivalence classes} among tilings  and prove that there exist exactly two such classes among soft tilings which share the full symmetry group of (e2). We also prove that if we only require tetrahedral symmetry for the cells but also prescribe the presence of at least one planar face, then there exist exactly four second-order classes of soft tilings  which are first order equivalents of (e2). Additionally, we construct a one-parameter family of tilings bridging standard and non-standard soft tilings, explicitly including the classic Kelvin’s foam structure as an intermediate configuration. This construction highlights that both the soft cells themselves and the geometric methods employed in their generation provide valuable insights into the structural principles underlying natural forms. We also present the soft tiling induced by the gyroid structure. 
\end{abstract}
\maketitle

\pagebreak
\tableofcontents
\pagebreak
\section{Introduction}

\subsection{Motivation and background}
In a recent article \cite{softcells1} we  introduced a new class of shapes, called \textit{soft cells} that fill space as  \emph{soft tilings} without gaps and overlaps while having no sharp corners at all. Soft tilings, like polyhedral tilings,  have a combinatorial structure
defined by the adjacency of vertices, edges and faces. However, it is important to note that edges need not be straight and faces need not be planar. We also defined a class of tilings called \emph{polyhedric}  that includes, as disjoint subsets, both soft tilings and convex tilings, the latter filling space with convex polyhedra. A key element for the construction of soft tilings is the  \textit{edge bending (EB) algorithm} that preserves the combinatorial structure and the location of vertices,  maps a subset of convex tilings (to which we refer as \emph{Hamiltonian} tilings) onto
as subset of soft tilings, to which we will refer as \emph{standard} soft tilings. Hamiltonian tilings have the property that at every node, the dual of the vertex polyhedron contains a Hamiltonian circuit. Standard soft tilings have the property that at every node, the half-tangents of every incoming edge are collinear. 

Although the EB algorithm guarantees the existence of a standard soft tiling $M'$ that is combinatorially equivalent to a given Hamiltonian tiling $M$, it does not yield a complete set of explicit instructions for the construction of $M'$.
Here we introduce the Extended Edge Bending (EEB) algorithm, which, in principle, is capable of computing, for any given polyhedral tiling $M$, the entire set $\mathcal{M}(M)$ of soft tilings that share the combinatorial structure and vertex locations of $M$, while satisfying a prescribed symmetry group and other geometric constraints. Depending on the symmetry group and constraints, the set $\mathcal{M}(M)$ may be finite or infinite and we will compute examples for both cases. The two algorithms are closely related: the EB algorithm guarantees that given a Hamiltonian tilings as input, the EEB algorithm has at least one standard soft tiling as output. 

\textit{Triply periodic minimal surfaces} (TPMS)  are embedded minimal surfaces in $\mathbb{R}^3$ that are invariant under three linearly independent translations, yielding a periodic structure that spans the entire space. The fundamental domains of the translational lattice under which the surface is invariant are  unit cells in their own right. Typically, these unit cells are  chosen such that the entire surface can be generated by translating the portion within a single cell along three linearly independent lattice vectors. The shape  of the unit cell depend on the specific symmetry group of the surface, and the surface within the cell satisfies the minimal surface condition (zero mean curvature) and matches smoothly across opposite faces under the lattice translations. Unit cells are visually not reminiscent of tilings, so, at first sight they are not related to soft cells either. On the other hand, one can construct Voronoi partitions of TPMS \cite{Schoen, grason} which yield  soft tilings; we will discuss such constructions related to the Schwarz P and Schwarz D surfaces \cite{Schwarz, Schoen} in Section \ref{sec:app}.  We will  show that the EEB algorithm not only  generates the soft tilings associated with these surfaces but it also computes families of soft tilings connecting them.  TPMS and their unit cells are critical in crystallographic and materials applications, as they define the repeating geometry and influence mechanical and transport properties when such surfaces are used as microstructure templates for materials \cite{grason, Han}.

Another interesting natural example is the \textit{Kelvin dry foam}. The corresponding cell, known as the \textit{Kelvin cell}, is a space-filling polyhedic cell proposed by Lord Kelvin in 1887  as a solution to the problem of partitioning space into cells of equal volume with minimal surface area \cite{Kelvin}. It is based on a truncated octahedron—a convex polyhedron with 6 square faces and 8 regular hexagonal faces. In order to satisfy Plateau’s laws for foams, Kelvin’s original construction slightly distorts the regular truncated octahedron into a polyhedric shape with curved faces to reduce the total interfacial area while still filling space without gaps.  The Kelvin cell approximates the minimal-area solution for discrete equal-volume polyhedric cells in a foam, whereas TPMS are smooth surfaces with zero mean curvature everywhere.

Here, we will describe and use the EEB algorithm to find non-standard soft tilings which are associated with triply periodic minimal surfaces, including the Schwarz P and Schwarz D surfaces and show that they connect soft tilings with the geometry of optimal dry foams \cite{Kelvin}.

\subsection{Basic notions and the main result}

Similarly to the EB algorithm, the output of the EEB algorithm does not fully characterize a soft tiling but only computes both half-tangents for each edge. The remaining features of the shape defined by the curves carrying edges and the surfaces carrying faces can be determined by additional considerations. Motivated by this common feature of the EB and EEB algorithms, below we introduce a classification scheme for polyhedric tilings:

\begin{defn}\label{def:levels}
Let $M$ be a polyhedric tiling. Then, the zeroth-order description of $M$ is the combinatorial structure of $M$. The $i$-th order description ($i=1,2,3,4$) of $M$  contains the $(i-1)$th order description and the following additional features:
\begin{itemize}
\item $i=1$:  the location of nodes, 
\item $i=2$:  the unit vectors of edge half-tangents, 
\item $i=3$:  the shape of edges, 
\item $i=4$:  the shape of faces.
\end{itemize}

If two tilings $M$ and $M'$ agree up to order $k$ then we say that they are elements of a $k$-th order
equivalence class  of tilings. The $i$-th order description is characterized by the symmetry group $\Gamma _i$ with fundamental domain $f_i$ and we have $\Gamma _i \leq \Gamma _j$ if $i \geq j$. 
\end{defn}

\begin{figure}[ht!]
\begin{center}
\includegraphics[width=\columnwidth]{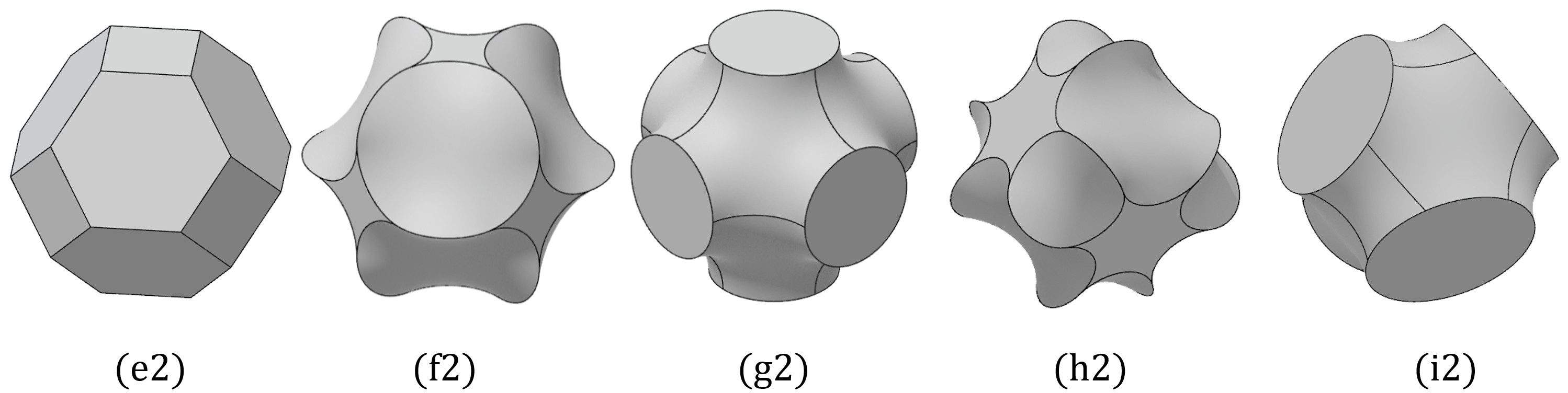}
\caption{The DV-$bcc$ cell (called the (e2) cell in \cite{softcells1}) and its four soft versions}\label{fig:1}
\end{center}
\end{figure}

Definition \ref{def:levels} naturally leads to a concise and practical new definition of softness:
\begin{defn}\label{def:soft}
Let $M$ be a polyhedric tiling with smooth edges and faces.
A cell of $M$ is \textit{soft}  if  each
node has at least two half-tangents unit vectors $\mathbf{u_1},\mathbf{u_2}$ 
such that $\mathbf{u_1}\cdot\mathbf{u_2}=-1$.
A tiling is \textit{soft} if each cell in the tiling is soft.
\end{defn}
Since softness depends on half-tangents, it is related to the second-order description of the tiling and it is independent of
any higher order features. As a consequence, both the EB and the EEB algorithms are of second-order, operating on half-tangents, so they do not
determine the shape of cells to higher order. If we want to construct a soft tiling explicitly, we need to specify the shape of its edges and faces, which are respectively included in the third- and fourth-order descriptions. Therefore, to obtain illustrative examples of
second-order soft equivalence classes, we will provide the full fourth-order description by
imposing  that edges are circular arcs with minimal curvature and faces are minimal surfaces.

In this paper we illustrate the EEB algorithm by using the monohedral Dirichlet-Voronoi tiling on the $bcc$ lattice as input. In \cite{softcells1} we called this the (e2) tiling and, using the EB algorithm, developed from (e2) the soft, standard (f2) tiling.(Figure \ref{fig:1}, first two cells from left).
Using the concepts in Definitions \ref{def:levels} and \ref{def:soft} we can now state our two main results for the (e2) tiling:
\begin{thm}\label{thm:0}
In the first-order equivalence class containing the (e2) tiling there exist exactly two second-order equivalence classes of soft tilings which share the full symmetry group of (e2). Figure \ref{fig:1} shows the soft cells (f2) and (g2) which respectively belong to these two classes.
\end{thm}

\begin{thm}\label{thm:1}
In the first-order equivalence class containing the (e2) tiling there exist exactly four second-order equivalence classes of soft tilings the cells of which have at least the symmetry group of the regular tetrahedron and have at least one planar face. Figure \ref{fig:1} shows the soft cells (f2), (g2),(h2) and(i2) which respectively belong to these four classes.
\end{thm}

 As we can observe, two of these cells are standard ((f2) and (h2)), the other two ((g2) and (i2)) are non-standard. In section \ref{sec:app} we will prove that the second-order equivalence classes containing (g2) and (i2) also contain, respectively, the unit cells of the Schwarz P and Schwarz D surfaces, serving as models of material microstructure, called \emph{mesoatoms} \cite{grason, Han}. 

If we drop the requirement that at least one face should be planar then the EEB algorithm generates a 1-parameter family of soft cells connecting the Schwarz D and Schwarz P unit cells.
If we drop the requirement of softness and only require that the quadrangular faces of the (e2) cell remain planar then the EEB algorithm computes a one-parameter family of spacefilling cells that,in addition to the (e2), (f2) and (h2) cells also includes the \emph{Kelvin cell}, serving as  the fundamental model of optimal dry foams \cite{Kelvin}.This demonstrates that soft cells are significant not only as final geometric forms but also in terms of their underlying geometric genesis which may provide valuable insights into modeling structures in natural phenomena.

We will prove Theorems \ref{thm:0} and \ref{thm:1} in two steps. We first introduce the EEB algorithm in Section \ref{sec:eeb}.
Then, in Section \ref{sec:e2} we apply the EEB algorithm to the (e2) tiling and compute all solutions
under the symmetry and geometric constraints stated in the Theorems. The list of all solutions in Table \ref{tab:allsolutions} completes the proofs which is illustrated in Figure \ref{fig:3}.
After proving Theorems  \ref{thm:0} and \ref{thm:1}, in Section \ref{sec:app}  we will discuss two applications: Schwarz minimal surfaces
and the Kelvin foam.

\section{The EEB algorithm}\label{sec:eeb}

We only describe the EEB algorithm for monohedric, isogonal tilings $M$ where all cells
and all nodes are identical, but all steps can be generalized to the case of multiple cells and nodes.
\begin{defn}\label{algorithm_input}
We assume that the polyhedric tiling  $M$ is defined to first order, so the symmetry group $\Gamma _1$ is known and we also know
the symmetry group $\Gamma _2 \leq \Gamma_1$ associated with the second order description. 
We assume that at each node  $N$ cells and $K$ edges meet, so
we have $K$ unit vectors as half-tangents, which we call the \emph{nodal set} of $M$.
From the nodal set we can construct
$N$ different subsets containing half-tangents, belonging to the $N$ cells.
We will refer to these sets of vectors as the $N$  \emph{vertex sets} of $M$
and we denote the size of the vertex sets by $v_i$, $i=1,2, \dots N$.
\end{defn}

\begin{algo}\label{def:EEB}
Using the notions in Definition \ref{algorithm_input}, the steps of the EEB algorithm are the following:

\begin{enumerate}
\item We identify the fundamental domain $f_2$ of $\Gamma _2$ in the second order description, i.e.
we identify  a maximal set of unit half tangents $\mathbf{u}_i$, $i=1,2, \dots f$ which are not related by any
transformation of the group $\Gamma _2$.
\item Using the set $\mathbf{u}_i$, $i=1,2, \dots f$ as inputs, we apply symmetry transformations
in $\Gamma _2$ to construct the nodal set $\mathbf{u}_i$, $i=1,2, \dots n$, $n\geq f$ of $M$.
\item We pick  a pair of half tangents $\mathbf{u}_i, \mathbf{u}_j$ from the nodal set and 
start writing a list:
\begin{equation*}\label{eq:core}
\mathbf{u}_i\mathbf{u}_j  =  -1
\end{equation*} 
We continue picking pairs and add the corresponding equation to list if the equation
does not agree with any previous equation and does not contradict any previous equation.
We continue this process until we have at least one equation in every vertex set.
At this point, we denote the number of equations by $E$ and we call this system of equations  
a  \emph{complete set of softening equations}
 of $M$. The same tiling may have several complete sets of softening equations.
\item We let the vectors $\mathbf{u}_i$, $i=1,2, \dots f$
of the fundamental domain  independently run over the boundary of unit sphere. (We may regard the $f$ unit sphere boundaries
as a configuration space or \emph{morphospace} \cite{budd}, containing all possible spacefilling cells of the first order family with given symmetry group.)
Since all vectors in the nodal set can be computed via transformation matrices from
the fundamental domain, this operation will turn the softening equations
into $E$ equations on the sphere  in 2 variables and we solve these systems.
In fact, the softening equations will be equivalent to eigenvalue problems for the transformation matrices.
\item If there are additional, prescribed constraints on the geometry of the cells (e.g. planar faces) then 
we also solve the corresponding equation systems.
\item We combine the solutions guaranteeing soft geometry with those
guaranteeing additional constraints. 
\end{enumerate}
\end{algo}
\begin{rem}
The existence of a  solution of any of these systems is a necessary (but not sufficient)
condition for the existence of a soft tiling $M'$ which is equivalent to first order
to $M$, has symmetry group $\Gamma_2$ and obeys the prescribed additional geometric constraints. 
Symmetry is not guaranteed since
we may have been able to compute the nodal set from the fundamental domain by applying only a subset of $\Gamma _2$.
\end{rem}

\section{The (e2) cell and the proof of Theorem \ref{thm:1}}\label{sec:e2}

In  this section we discuss the application of the EEB algorithm to the  Dirichlet-Voronoi cell on the $bcc$ lattice (the (e2) cell), shown in Figure \ref{fig:2} (see also Figure \ref{fig:1}, left image).
Our goal is to prove Theorems \ref{thm:0} and \ref{thm:1}. We will do the proof in steps numbered as subsections and the numbering of these subsections corresponds \emph{exactly} to the numbering of the steps in Algorithm  \ref{def:EEB}. For steps 1, 2 and 3 the proofs for both Theorems agree. For  steps 4, 5 and 6 we will discuss
the two proofs separately. The proofs are concluded after Table \ref{tab:allsolutions}.

\begin{proof}

The first-order description of the cell is provided in Table \ref{tab:dv_bcc}, defined as the (e2) cell in \cite{softcells1}. While Figure \ref{fig:2} shows the polyhedral cell defined to fourth order, we only use the first-order information contained in Table \ref{tab:dv_bcc}.

\begin{figure}[ht!]
\begin{center}
\includegraphics[width=0.5\columnwidth]{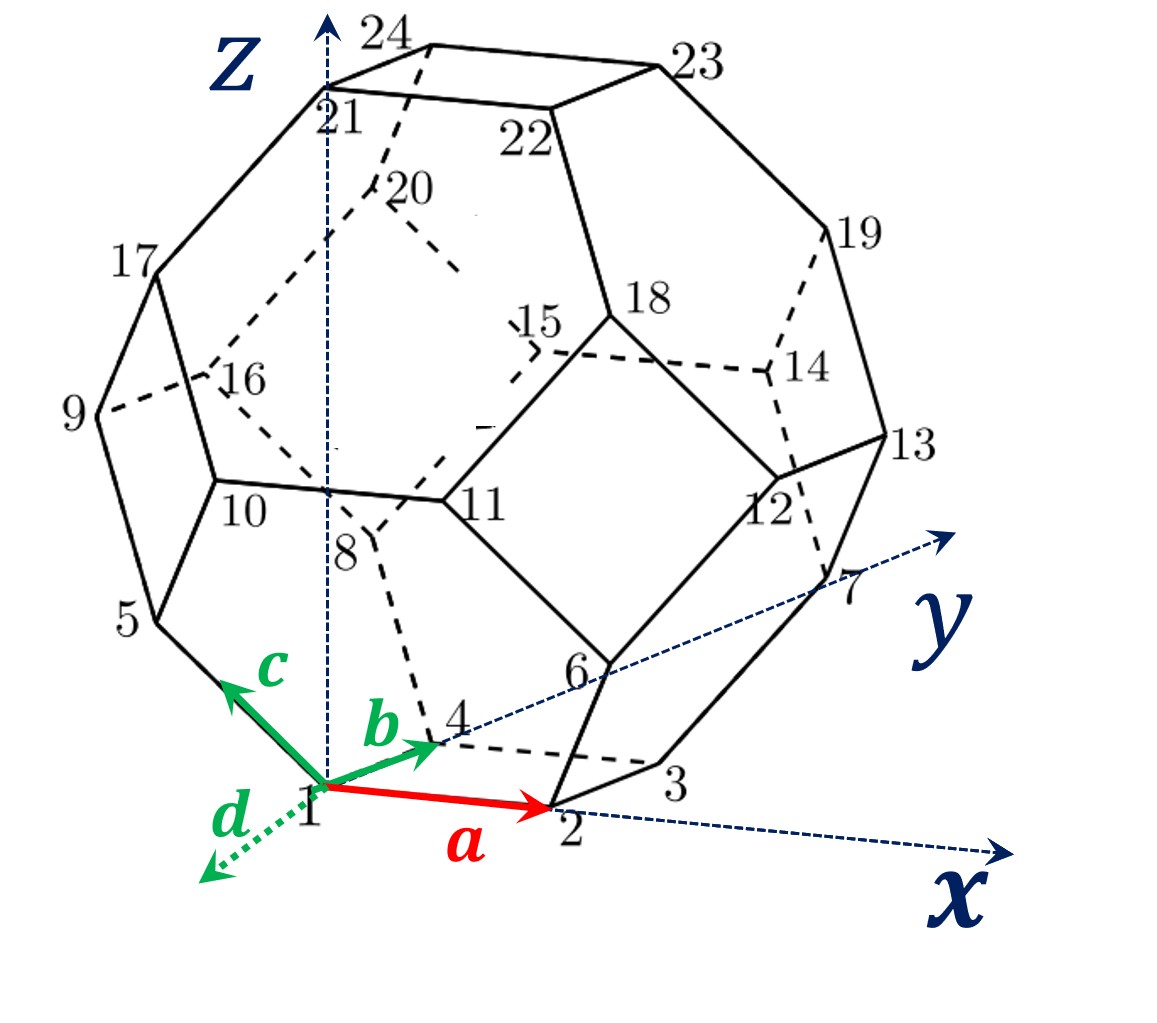}
\caption{The DV-$bcc$ cell}\label{fig:2}
\end{center}
\end{figure}
\begin{table}[ht!]
    \centering
    \begin{tabular}{|c||c|c|c|| c || c | c | c |}
         \hline
				  Node & $x$ & $y$  & z &   Node & $x$ & $y$  & z \\
         \hline
         \hline
         1 & 0 & 0 & 0 & 13 & 2 & 1 & $2/\sqrt{2}$ \\  
				\hline
				  2 & 1 & 0 & 0 & 14 & 1 & 2 & $2/\sqrt{2}$ \\  
				\hline
				 3 &  1& 1 & 0 & 15 & 0 & 2 & $2/\sqrt{2}$ \\  
				\hline
				  4 & 0 & 1 & 0 & 16 & -1 & 1 & $2/\sqrt{2}$ \\  
				\hline
				 5 & -1/2 & -1/2 & $1/\sqrt{2}$ & 17 & -1/2 & -1/2 & $3/\sqrt{2}$ \\  
				\hline
				  6 & 3/2 & -1/2 & $1/\sqrt{2}$ & 18 & 3/2 & -1/2 & $3/\sqrt{2}$ \\  
				\hline
				 7 & 3/2 & 3/2 & $1/\sqrt{2}$ & 19 & 3/2 & 3/2 & $3/\sqrt{2}$ \\  
				\hline
				  8 & -1/2 & 3/2 & $1/\sqrt{2}$ & 20 & -1/2 & 3/2 & $3/\sqrt{2}$ \\  
				\hline
				 9 & -1 & 0 & $2/\sqrt{2}$ & 21 & 0 & 0 & $4/\sqrt{2}$ \\  
				\hline
				  10 & 0 & -1 & $2/\sqrt{2}$ & 22 & 1 & 0 & $4/\sqrt{2}$ \\  
				\hline
				 11 & 1 & -1 & $2/\sqrt{2}$ & 23 & 1 & 1 & $4/\sqrt{2}$ \\  
				\hline
				  12 & 2 & 0 & $2/\sqrt{2}$ & 24 & 0 & 1 & $4/\sqrt{2}$ \\  
				\hline
							\end{tabular}
    \caption{First order description of the (e2) cell}
    \label{tab:dv_bcc}
\end{table}

\subsection{Symmetries and the fundamental domain.}
The \emph{space group}  of the (e2) tiling is  $ \Gamma _1=Im3m$  (listed as \# 229 in \cite{symmetry}), containing the symmetry group of the regular octahedron as a \emph{point group} of order 48.
In Theorem \ref{thm:0}, for the second order description we require the same symmetry groups.
In Theorem \ref{thm:1}, for the second order description we require that the soft tiling should have the $\Gamma _2 = Pn3m < \Gamma _1$ space group (listed as \# 224 in \cite{symmetry}), containing the symmetry group of the regular tetrahedron  as a point group of
order 24. In both cases (Theorems \ref{thm:0} and \ref{thm:1}) the fundamental domain $f_2$ of the second order description is a single unit vector (i.e. we have $f=1$) and we pick  $\mathbf{a}$
(in the polyhedral tiling  $\mathbf{a}=(1,0,0)^T$). So we have
$\mathbf{u}_1\equiv \mathbf{a}.$
 
The nodal set of this tiling consists of 4 vectors (so we have $n=4$)  which we show in Figure \ref{fig:2} at vertex 1,  denoted by $\mathbf{a}, \mathbf{b}, \mathbf{c}, \mathbf{d}$, respectively,
so we have
$
\mathbf{u}_2\equiv \mathbf{b}, \quad \mathbf{u}_3\equiv \mathbf{c}, \quad \mathbf{u}_4\equiv \mathbf{d}.
$
These vectors  can be obtained from the vector $\mathbf{a}$ via the following linear transformations:
\begin{equation}
\label{eq:transformations}
\mathbf{b}  =  T_{b} \mathbf{a}, \quad
\mathbf{c}  = T_{c} \mathbf{a}, \quad
\mathbf{d}  = T_{d,i} \mathbf{a}, \quad (i=1,2)
\end{equation}
where
\begin{equation}\label{eq:matrices_bc}
T_{b}=\left(
\begin{array}{ccc}
0 & 1& 0 \\
1 & 0 & 0 \\
0 & 0 & -1
   \end{array}
\right), \quad
T_{c}=
\left(
\begin{array}{ccc}
-\frac{1}{2} & -\frac{1}{2} & \frac{1}{\sqrt{2}}\\
-\frac{1}{2} & -\frac{1}{2} & -\frac{1}{\sqrt{2}}\\
\frac{1}{\sqrt{2}} & -\frac{1}{\sqrt{2}} & 0 \\
\end{array}
\right)
\end{equation}
\begin{equation}\label{eq:matrices_d}
T_{d,1}=\left(
\begin{array}{ccc}
-\frac{1}{2} & -\frac{1}{2} & +\frac{1}{\sqrt{2}}\\
-\frac{1}{2} & -\frac{1}{2} & -\frac{1}{\sqrt{2}}\\
-\frac{1}{\sqrt{2}} & +\frac{1}{\sqrt{2}} & 0 \\
\end{array}
\right), \quad
T_{d,2}=\left(
\begin{array}{ccc}
-\frac{1}{2} & -\frac{1}{2} & -\frac{1}{\sqrt{2}}\\
-\frac{1}{2} & -\frac{1}{2} & +\frac{1}{\sqrt{2}}\\
-\frac{1}{\sqrt{2}} & +\frac{1}{\sqrt{2}} & 0 \\
\end{array}
\right).
\end{equation}
The transformations $T_b,T_c$ apply for both symmetry groups (and thus both Theorems). The transformation producing the vector $\mathbf{d}$ is different in
the two cases, $T_{d,1}$ applies for octahedral symmetry, in the proof of Theorem \ref{thm:0} and $T_{d,2}$ applies for tetrahedral symmetry, in the proof of Theorem \ref{thm:1}. 
\subsection{Nodal set and vertex sets}
Since we are dealing with a primitive Voronoi tiling,  four cells meet at a node so we have $N=4$. For the same reason, the size of all vertex sets is equal, we have $v_i=3$, $i=1,2,3,4$.
The vectors  $(\mathbf{a},\mathbf{b},\mathbf{c})$ compose the vertex set of the cell visible in the figure.  There are three other
cells meeting at this vertex and their respective vertex sets are $(\mathbf{a},\mathbf{b},\mathbf{d})$, $(\mathbf{a},\mathbf{c},\mathbf{d})$ and $(\mathbf{b},\mathbf{c},\mathbf{d})$.
\subsection{Complete sets of softening equations}
For this system we can write three different complete sets of softening equations:
\begin{eqnarray} \label{abcd}
\mathbf{a}\mathbf{b} = -1, & \mathbf{c}\mathbf{d} = -1 \\
\label{acbd}
\mathbf{a}\mathbf{c} = -1, & \mathbf{b}\mathbf{d} = -1 \\
\label{adbc}
\mathbf{b}\mathbf{c} = -1, & \mathbf{a}\mathbf{d} = -1.
\end{eqnarray}

\subsection{Solution of the softening equations}
Now we make step (4) of the algorithm and let the fundamental domain of the 2nd order description (the unit vector $\mathbf{a}$) run over the sphere.
\subsubsection{Octahedral symmetry: solutions of the equations in the proof of Theorem \ref{thm:0}}
First we solve system (\ref{abcd}). The first equation yields the great circle $g_{ab}$ with unit normal
\begin{eqnarray}\label{uab}
\mathbf{u}_{ab} & = & \left(  
\frac{1}{\sqrt{2}},  \frac{1}{\sqrt{2}}, 0 \right)^T,
\end{eqnarray}
while the second equation has an isolated solution at 
\begin{eqnarray}\label{c1d1}
\mathbf{a}_{cd} & = & \left(  
\frac{1}{\sqrt{2}},  -\frac{1}{\sqrt{2}}, 0 \right)^T.
\end{eqnarray}
Next we solve system (\ref{acbd}). The first equation yields the great circle $g_{ac}$ with unit normal
\begin{eqnarray}\label{uac}
\mathbf{u}_{ac} & = & \left(  
\frac{1}{{2}},  -\frac{1}{{2}}, \frac{1}{\sqrt{2}} \right)^T, 
\end{eqnarray}
while the second equation has an isolated solution at 
\begin{eqnarray}\label{b1d1}
\mathbf{a}_{bd} & = & \left(  
\frac{1}{\sqrt{2}},  \frac{1}{\sqrt{2}}, 0 \right)^T.
\end{eqnarray}
Next we solve system (\ref{adbc}). The first equation yields the great circle $g_{bc}$ with unit normal
\begin{eqnarray}\label{ubc}
\mathbf{u}_{bc} & = & \left(  
-\frac{1}{{2}},  \frac{1}{{2}}, \frac{1}{\sqrt{2}} \right)^T, 
\end{eqnarray}
while the second equation has an isolated solution at 
\begin{eqnarray}\label{a1d1}
\mathbf{a}_{ad} =\mathbf{a}_{bd} & = & \left(  
\frac{1}{\sqrt{2}},  \frac{1}{\sqrt{2}}, 0 \right)^T.
\end{eqnarray}

\subsubsection{Tetrahedral symmetry: solutions of the equations in the proof of Theorem \ref{thm:1}}
If we solve the three equation systems (\ref{abcd}),(\ref{acbd}) and (\ref{adbc}) for the unit vector $\mathbf{a}$,
the solutions are  great circles
which denote respectively by  $g_{abcd} \equiv g_{ab}, g_{acbd} \equiv g_{ac}$ and $ g_{adbc}\equiv g_{bc}$
the respective unit normals given in (\ref{uab}), (\ref{uac}) and (\ref{ubc}).  
We can observe that $g_{acbd}$ and $g_{adbc}$ are related by symmetry with respect to the diagonal plane $x=y$,
so henceforth we will only compute solutions related to the former.

\subsection{Additional constraints:  planar faces}
In Theorem \ref{thm:0} there are no additional constraints, however, in Theorem \ref{thm:1} we added the constraint that at least one face should remain planar so this subsection is related only to the proof of Theorem \ref{thm:1}. 
For planar face constraints it follows that if a great circle solves the problem then it must
lie in the plane of the given face. We solve the corresponding
constraint equations to ensure that such a solution
actually exists.

The condition that the quadrangular face (1,2,3,4) should remain planar can be written  as
\begin{equation}\label{eq:quad}
\mathbf{u}_{quad}\mathbf{a} =  0, \quad \mathbf{u}_{quad}\mathbf{b}  =  0, 
\end{equation}
where $\mathbf{u}_{quad}=(0,0,1)^T$ is the unit normal of the quadrangular face.The
solution of the system (\ref{eq:quad}) is the great circle $g_{quad}$ with unit normal
$\mathbf{u}_{quad}.$ 

The condition that the hexagonal face
(1,2,6,10,11,5) should remain planar can be written as
\begin{equation}\label{eq:hex}
\mathbf{u}_{hex1}\mathbf{a}=0, \quad \mathbf{u}_{hex1}\mathbf{c}=0,
\end{equation}
where $\mathbf{u}_{hex1}=\left( 0, 2/\sqrt{6} , \sqrt{2}/\sqrt{6}  \right)^T$ is the unit normals of the hexagonal face.
An analogous condition for face (1,4,8,16,9,5) yields $\mathbf{u}_{hex2}=\left( 2/\sqrt{6},0 , \sqrt{2}/\sqrt{6}  \right)^T$. 
These solutions correspond to the great circles $g_{hex1}, g_{hex2}$ with respective unit normals
$\mathbf{u}_{hex1}$, $\mathbf{u}_{hex2}.$

\begin{figure}[ht!]
\begin{center}
\includegraphics[width=\columnwidth]{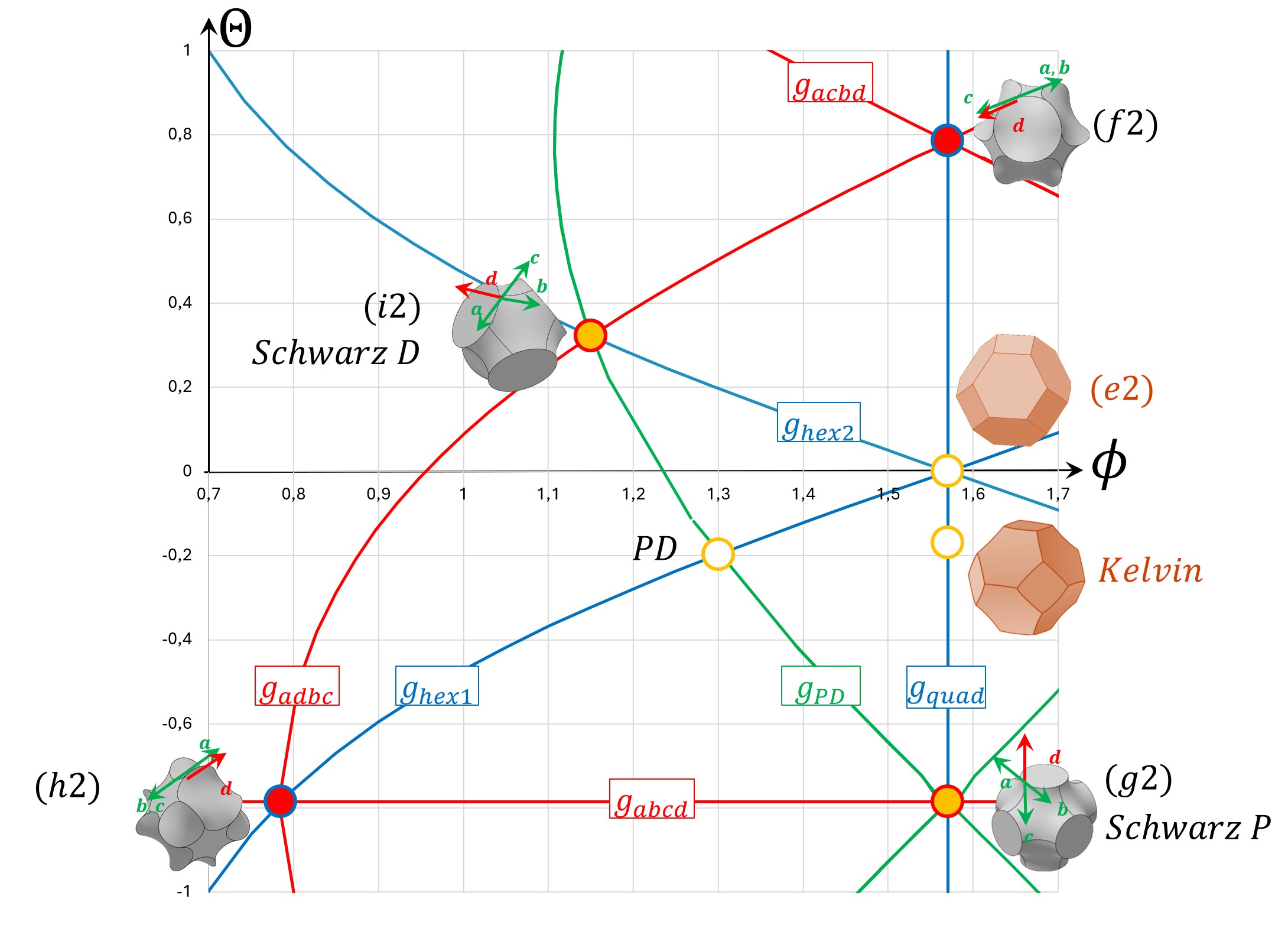}
\caption{All solutions of the softening equations for the polyhedral (e2) cell shown in Figure \ref{fig:2}, plotted
on the plane $(\phi, \theta)$ of Euler angles ($\theta=0,  \phi=0 $ coinciding respectively with the $x$ and $z$ axes of the Cartesian system shown on Figure \ref{fig:2}).
Red lines: great circles solving complete sets of softening equations. Blue lines: great circles solving constraint equations which keep faces planar. Green lines:great circles connection non-standard soft cells.
 Blue dots with red fill: standard soft cells identified by the intersection of two red lines and one blue line.
 Red dots with yellow fill: non-standard soft cells defined by the intersection of one blue and one red line.Yellow dots with white fill: non-soft cells.
For detailed data on cells (e2),(f2),(g2),(h2),(i2) see Table \ref{tab:allsolutions}. For data on the Kelvin cell and the $PD$ cell see Table \ref{tab:KelvinPD}.
For better visibility, nodal set $\mathbf{a},\mathbf{b}, \mathbf{c}, \mathbf{d}$ is shown on vertex 21 instead on vertex 1.  }\label{fig:3}
\end{center}
\end{figure}

\subsection{Combining solutions: isolated cells}
Now we can conclude the proofs of Theorems \ref{thm:0} and \ref{thm:1} and we discuss these separately.
\subsubsection{Octahedral symmetry: completing the proof of Theorem \ref{thm:0}}
As we can observe, all equation systems yield one great circle and one isolated point on that great circle.
This would provide three isolated solutions, however, the solutions of the systems (\ref{acbd}) and (\ref{adbc}) are identical,
so we obtain two isolated solutions, given in (\ref{c1d1}) and (\ref{b1d1}), corresponding to the soft cells
(f2) and (g2), respectively, listed in Table \ref{tab:allsolutions}.
Figure \ref{fig:3} illustrates, among others, these 2  soft cells along with  the  great circles
$g_{abcd}, g_{acbd}$ on the plane $(\phi, \theta)$ of the Euler angles.

\subsubsection{Tetrahedral symmetry: completing the proof of Theorem \ref{thm:1}}
We observe that all equation systems yield a one-parameter family of solutions, each family being equivalent to a great
circle on the unit sphere. To obtain \emph{isolated} solutions we regard suitable \emph{pairs} of these great circles.
We pick pairs in such a manner that for each pair one great circle should guarantee the softness, the other
should guarantee at least one planar face. We also consider that the great circles $g_{acbd}$ and $g_{adbc}$
are related by the $x \leftrightarrow y$ reflection symmetry, so it is sufficient to consider one of them
to obtain all solutions which are not related by a symmetry transformation. Using these considerations
we obtained the  four soft solutions (f2),(g2), (h2) and (i2), listed in Table \ref{tab:allsolutions}. The polyhedral cell (e2) is listed in the
first row of the table, as it can  be
obtained by combining the solutions for planar hexagonal and planar quadrangular faces.
In the table we not only identify the great circles the intersection of which determine the given cell
but also give the Cartesian coordinates and the corresponding Euler angles for the unit vector $a$.
From the latter the entire tiling can be reconstructed to second order by the action  of the space group $\Gamma_2=Pn3m$.

Figure \ref{fig:3} is illustrating both
the  four soft cells satisfying all constraints, but also the polyhedral cell $(e2)$ and the  great circles
$g_{abcd}, g_{acbd}, g_{hex1}, g_{hex2}$ and $g_{quad}$ on the plane $(\phi, \theta)$ of the Euler angles.

We note that the listed solutions are only defined to second order, i.e each soft cell 
that appears as an isolated point on the plane of Euler angles  represents
an infinite set of soft cells with identical vertex locations and identical half-tangents for the edges. The cells appearing in
Figures \ref{fig:1}, \ref{fig:3} are particular elements of this infinite sets. We picked these elements
by using circular arcs as edges and minimal surfaces as faces.

\begin{table}[ht!]
    \centering
    \begin{tabular}{||c||c|c| c||c|c|c|c|c||c|}
         \hline
				  Name & \multicolumn{3}{c||}{great circles} & $a_x$ & $a_y$  & $a_z$ & $\phi$ & $\theta$ & $\sigma$ \\
         \hline
         \hline
			 	(e2) & $g_{hex}$ & $g_{quad}$ &  & 1 &  0  & 0 & $\pi/2$   &  0 & 0 \\
				\hline
				\hline
        (f2) & $g_{adbc}$ & $g_{quad}$& $g_{acbd}$  & $1/\sqrt{2}$ & $1/\sqrt{2}$  & 0 & $\pi/2$   &  $\pi/4$  & 0.331 \\
				\hline
				(g2) & $g_{adbc}$ & $g_{quad}$ &  & $1/\sqrt{2}$ & $-1/\sqrt{2}$  & 0 & $\pi/2$   &  $-\pi/4$ &  0.333 \\
				\hline
				(h2) & $g_{adbc}$ & $g_{hex1}$ & $g_{acbd}$ & $1/2$ & $-1/2$  & $1/\sqrt{2}$ & $\pi/4$   &  $-\pi/4$ & 0.464  \\
				\hline
				(i2) & $g_{adbc}$ & $g_{hex2}$ & & $\sqrt{3}/2$ & $\sqrt{3}/6$  & $1/\sqrt{6}$ & $\cos^{-1}(1/\sqrt{6})$   & $\tan^{-1} (1/3)$  & 0.474 \\
				\hline
							\end{tabular}
    \caption{The polyhedral (e2) cell and all 4 first order equivalent soft cells with at least tetrahedral symmetry and at least one planar face, given with
		the $[x,y,z]$ coordinates of the unit vector $\mathbf{a}$ (cf. Figure \ref{fig:2}) and with Euler angles, where $\theta=0,  \phi=0 $ coincide respectively with the $x$ and $z$ axes of the Cartesian system shown on Figure \ref{fig:2}. Last column: softness value $0 \leq \sigma \leq 1$, as defined in \cite{softcells1}.}
    \label{tab:allsolutions}
\end{table}
\end{proof}

\section{Applications}\label{sec:app}
\subsection{Schwarz minimal surfaces}
As previously introduced, Triply Periodic Minimal Surfaces (TPMS) are smooth, continuous, minimal surfaces (i.e. with zero mean curvature) characterized by three independent, periodic directions in three-dimensional space.
Beyond their unique geometrical role \cite{Mackay}, highly relevant as models in material science \cite{grason, Han}.
While the primary geometric interpretation of a TPMS is that of an \emph{interface}, the same object can also be interpreted as a \emph{tiling}.
This interpretation is certainly not new. While many aspects have been described both from the geometrical \cite{Schoen} and  physical \cite{grason} perspective,
we are not aware of a clear definition connecting TPMS to tilings. Hence, so our first goal is to define this connection.

\begin{figure}[h!]
\begin{center}
\includegraphics[width=\columnwidth]{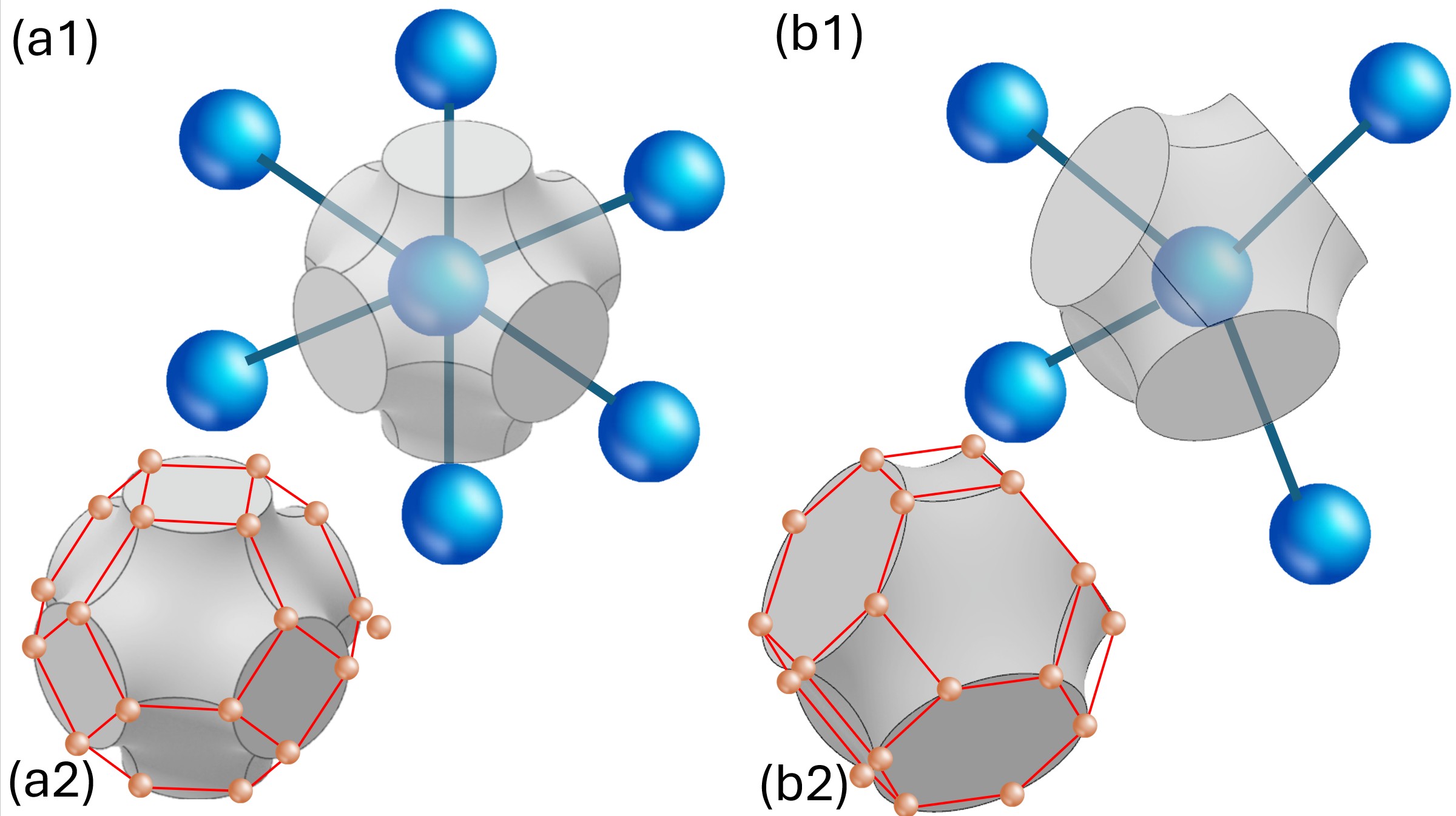}
\caption{Discrete structures on triply periodic minimal surfaces. Upper row, (a1)-(b1): Unit cells and skeletal graphs. Bottom row, (a2)-(b2): Unit cells carrying the vertices of the skew polyhedron $\{6|4,4\}$.
Left column, (a1)-(a2): Schwarz P surface. Right column, (b1)-(b2): Schwarz D surface.}\label{fig:4}
\end{center}
\end{figure}

TPMS generate a special binary partition of 3D space: each partition is a \emph{labyrinth} formed by a tubular system
with infinitely many branching points \cite{Schoen}. Hence, a given TPMS gives rise to two tubular labyrinths with branching points.
Both tubular labyrinths can be collapsed onto a respective \emph{skeletal graph} the nodes of which correspond to branching points
(see Figure \ref{fig:4}, panels (a1),(b1)), so TPMS are associated to a pair of infinite skeletal graphs.
Since  TPMS are smooth, so the cross section of the tubes is also a smooth curve. This implies a simple
\begin{obs}\label{ob:schwarz}
The Voronoi partition of a labyrinth of a TPMS labyrinth with respect to the nodes of its skeletal graph results in a soft tiling.
We will call the cell of this tiling the Voronoi cell associated with the TPMS.
If the two labyrinths are identical and the skeletal graph has identical edges and vertices then this tiling is monohedric
and the Voronoi cell of the tiling is a monohedric soft cell. Monohedric Voronoi tilings ahve been used
as physical models of materials structure, and they were called \emph{mesoatoms} \cite{grason}.
\end{obs}
The first TPMS was identified by Riemann \cite{Riemann} and independently  by Schwarz \cite{Schwarz} who also found another example;
these two surfaces are dubbed as the Schwarz D ("`diamond"') and the Schwarz P ("`primitive"')
surface, respectively. Panels (a1) and (b1) of Figure \ref{fig:4} show a Voronoi cell of these surfaces along with a portion of their skeletal graphs.
For both the Schwarz P and the Schwarz D surface, the dual labyrinth is identical to the one shown in the figure and 
the skeletal graphs are isogonal, i.e. their nodes are identical. The degree $n$ of their nodes is characteristic of the symmetry of their Voronoi cells,
as it defines the number of neighbor  cells in the same labyrinth.
In the case of the Schwarz P surface we have  a node of degree $n=6$ with cubic symmetry and in the case of the Schwarz D surface we have $n=4$
with tetrahedral symmetry. Remarkably, these surfaces are related to the soft tilings we demonstrated in Section \ref{sec:eeb}:
\begin{prop}\label{prop:schwarz}
The Voronoi cells of the Schwarz P and Schwarz D minimal surfaces are equivalent to second order to the soft cells (g2) and (i2), respectively.
\end{prop}

\begin{proof}
Here we regard the soft tilings defined by two TPMS: the Schwarz P and Schwarz D surfaces, illustrated in Figure \ref{fig:4}. According to Schoen \cite{Schoen},
the unit cell of the Schwarz P surface has octahedral symmetry (see also \cite{gandy}) and the unit cell of the Schwarz D surface has tetrahedral symmetry
and both are equivalent to the (e2) tiling to first order: they carry the vertices of the regular map  $\{6,4 |4\}$ \cite{Coxeter} which is, to third order,
composed of the truncated octahedra (see panels (a2) and (b2) of Figure \ref{fig:4}). Based on Observation \ref{ob:schwarz} we also
know that both Schwarz Voronoi cells have $n$ smooth, planar faces. According to Theorem \ref{thm:1}, there are 4 soft tilings 
which are first order equivalents of the (e2) tiling and also carry planar faces. Two of these tilings has a unit cell where the planar faces are non-smooth: these
are the standard soft cells (f2) and (h2). The only remaining two cells are (g2) and (i2) so the two Schwarz cells have to be equivalent to these cells to second order.
\end{proof}

\subsection{The Kelvin cell}
\begin{figure}[h!]
\begin{center}
\includegraphics[width=0.6\columnwidth]{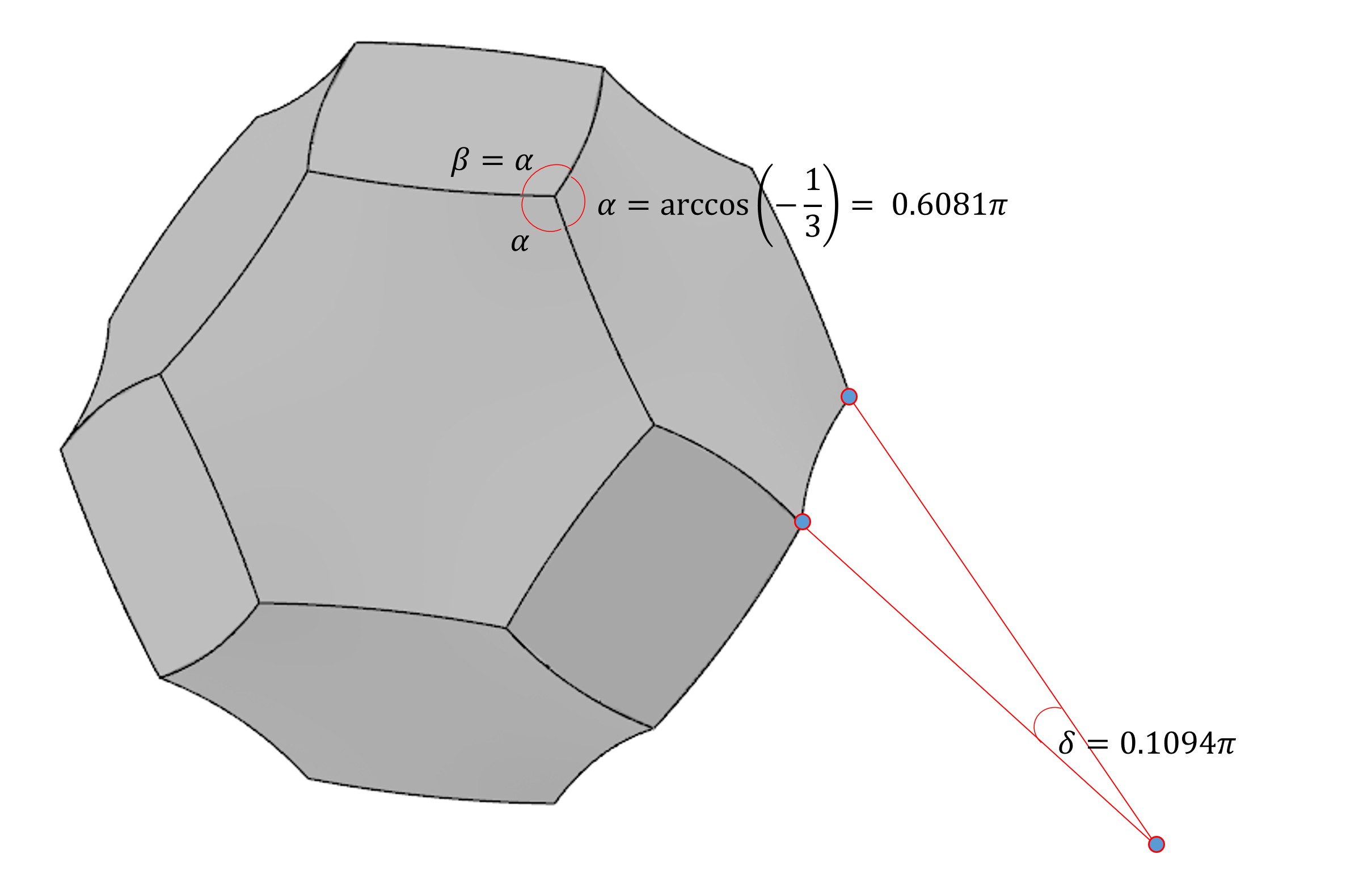}
\caption{The Kelvin Cell }\label{fig:5}
\end{center}
\end{figure}

A notable, historic example for curved tilings is the geometric model of dry foams where the tiling has to obey Plateau's Laws \cite{Plateau}.
While the structure of foams is strongly reminiscent of polyhedral tilings, there is no space-filling polyhedron 
meeting this requirement. William Thomson (Lord Kelvin), when faced with this dilemma, proposed to \textit{slightly modify} the truncated octahedron by deforming its edges \cite{Kelvin}.
The result is a monohedric tiling  that became known as the \emph{Kelvin  Foam} and its cell as the \emph{Kelvin Cell} (see Figure \ref{fig:5}). The latter is, to this date, is believed to have the smallest surface area among solids that tile space 
 \cite{Weaire2007, Bitsche_2005, daxner_2006}.

The Kelvin foam is identical to the (e2) tiling (see Figure \ref{fig:2}) to first order and it is invariant under the $\Gamma _1 = Im3m$ space group. 
 Similarly to the previously discussed soft cells, the fundamental domain is 
represented by the vector $\mathbf{a}$. Below we compute the coordinates of $\mathbf{a}$, both in the Cartesian $(x,y,z)$ system as well in Euler angles.

The geometric constraint equations defining the Kelvin cell to second order describe Plateau's law for the edges, prescribing that all edges meet at identical angles, yielding:
\begin{equation}\label{Kelvin1}
\mathbf{a}\mathbf{b} = \mathbf{a}\mathbf{c} = \mathbf{a}\mathbf{d} = \mathbf{b}\mathbf{c} = \mathbf{b}\mathbf{d} = \mathbf{c}\mathbf{d}.
\end{equation}
We find that from the (\ref{Kelvin1}) system the equation $\mathbf{a}\mathbf{c} = \mathbf{a}\mathbf{d}$ yields 
\begin{equation}\label{az}
a_z=0.
\end{equation}
Also, from (\ref{Kelvin1}) it follows that the 4 vectors point to the vertices of a regular tetrahedron, yielding
\begin{equation}\label{Kelvin2}
\mathbf{a}\mathbf{b} = \mathbf{a}\mathbf{c} = \mathbf{a}\mathbf{d} = \mathbf{b}\mathbf{c} = \mathbf{b}\mathbf{d} = \mathbf{c}\mathbf{d}=-1/3.
\end{equation}
If we pick $\mathbf{a}\mathbf{b} =-1/3$ from (\ref{Kelvin2}) and we substitute (\ref{az})
into this equation, we get the solution displayed in Table \ref{tab:KelvinPD}. This
also satisfies  (\ref{eq:quad}), the Kelvin cell has  $Im3m$ space symmetry, the tiling is
 invariant under the $z \to -z$ reflection, implying that the quadrangular faces of the cell are planar. 

We observe that the $g_{quad}$ great circle contains a one parameter family of spacefilling cells which are first order equivalents of the (e2)
cell and which connect the latter with the (f2) standard soft cell and the Kelvin cell.

\begin{table}[ht!]
    \centering
    \begin{tabular}{||c||c| c||c|c|c|c|c|}
         \hline
				  Name & \multicolumn{2}{c||}{great circles} & $a_x$ & $a_y$  & $a_z$ & $\phi$ & $\theta$ \\
         \hline
         \hline
			 	Kelvin & $g_{quad}$ &  & $\frac{1}{\sqrt{1+(2\sqrt{2}-3)^2}}$  & $\frac{2\sqrt{2}-3}{\sqrt{1+(2\sqrt{2}-3)^2}}$ &  0  &  $\frac{\pi}{2}$   &  $\tan^{-1}(2\sqrt{2}-3)$ \\
				\hline
        $PD$ & $g_{PD}$ & $g_{hex1}$&  $\frac{5}{\sqrt{28}}$ & $\frac{-1}{\sqrt{28}}$  & $\frac{\sqrt{2}}{\sqrt{28}}$ & $ \cos^{-1}\left(\frac{1}{\sqrt{14}}\right)$   &  $\tan^{-1}(-\frac{1}{5})$ \\
				\hline
							\end{tabular}
    \caption{The Kelvin cell and the $PD$ cell. See Figure \ref{fig:3} for the representation on the plane of the Euler angles.}
    \label{tab:KelvinPD}
\end{table}

%\end{proof}

\section{Open questions and summary}

\subsection{The gyroid cell}
The Schwarz P and D surfaces are related
by the continuous bending transformation described in 1853 by Ossian Bonnet \cite{Bonnet}.
They are members of a one-parameter family of surfaces, called the Bonnet family,
with free parameter $\alpha$, at the  values $\alpha=0$ (D)  and $\alpha=\pi/2$ (P).
Beyond the Schwarz P and D surfaces there one single member of this family
(at $\alpha \approx 0.663225$) which has no self-intersections: this TPMS, discovered by Alan Schoen \cite{Schoen}, is called the gyroid.

\begin{figure}[h!]
\begin{center}
\includegraphics[width=\columnwidth]{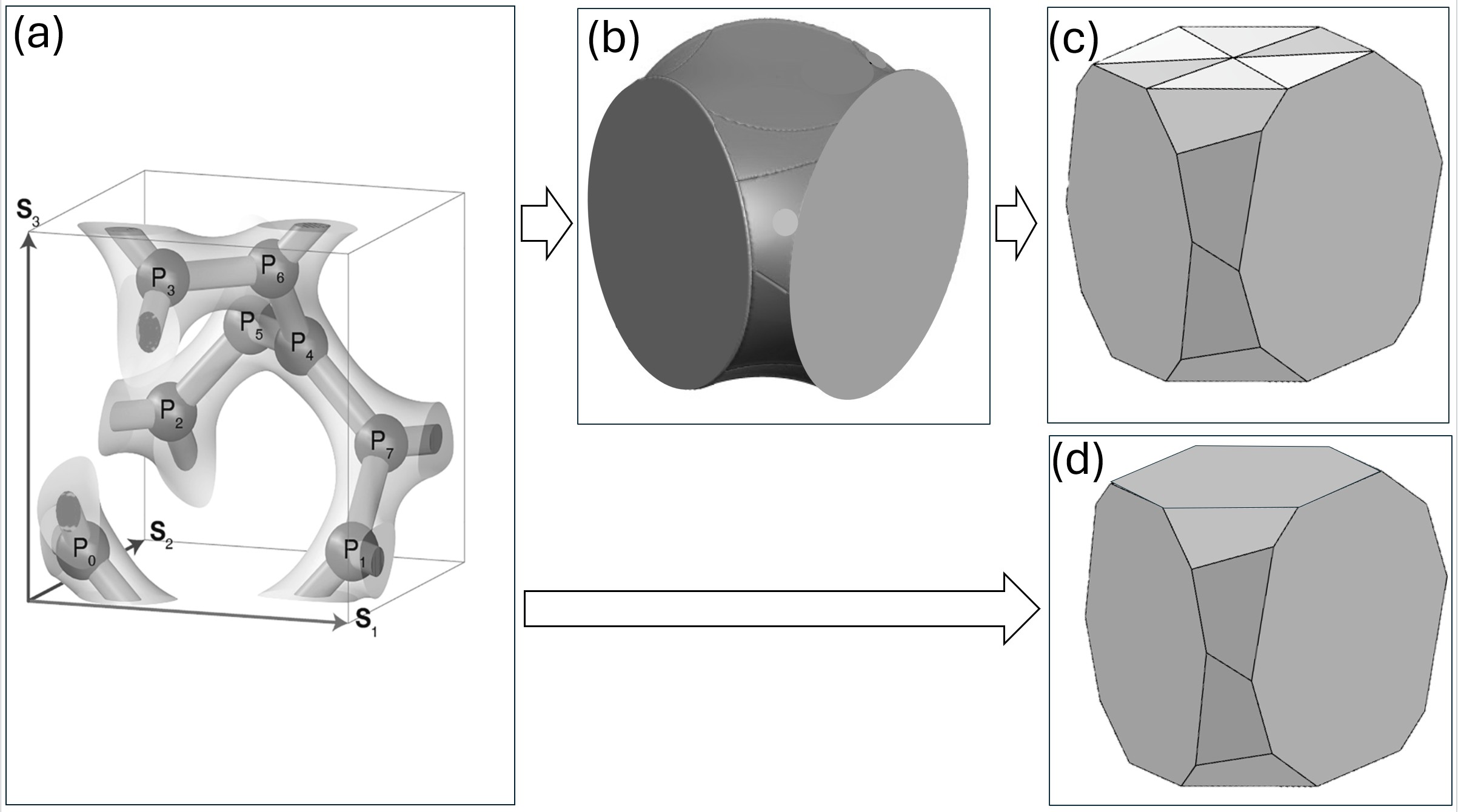}
\caption{The gyroid cell. (a) One labyrinth of the gyroid with $n=3$ order nodes $P_i$, $i=0,1,2,3,4,5,6,7$ (see \cite{Park}, Figure 1e).
 (b) Soft, space-filling $G$ cell, obtained
by Voronoi decomposition of the gyroid labyrinth \cite{grason}. (c) Non-convex, space-filling monohedral cell obtained by first order approximation of the $G$ cell.
(d) Convex, space-filling monohedral cell $L_2V_{17}$, obtained by Voronoi decomposition of 3D space using the nodes $P_i$ \cite{Schoen}.}\label{fig:gyroid}
\end{center}
\end{figure}

While the gyroid is included in the same Bonnet family as the Schwarz P and D surfaces, its
geometry is radically different. While the gyroid also gives rise to two identical labyrinths (differing only in handedness), however,
the nodes of the  skeletal graphs are
of degree $n=3$ and in a unit cube there are 8 such nodes (see panel (a) of Figure \ref{fig:gyroid}, based on Figure 1e in \cite{Park}).
According to Schoen \cite{Schoen}, the space group of the gyroid surface is the $I4_132$ group
(\# 214 in the list provided in \cite{symmetry}).

Panel (a) of Figure \ref{fig:gyroid} shows the qualitative picture of one labyrinth of the gyroid. Using Observation \ref{ob:schwarz}
one can obtain the soft, monohedric G cell in panel (b). We computed the softness value $\sigma$ for this cell and found $\sigma=0.576$ which
is the highest value computed so far for any monohedric soft cell.  Subsequently, by removing all edges and all faces, we took the first-order approximation 
of this soft cell and obtained the \emph{non-convex}, spacefilling polyhedron, shown in panel (c). (Recall, that in the case
of the Schwarz P and Schwarz D surfaces this operation results in the convex (e2) cell.) In the next step, we used  the nodes $P_i$ of the labyrinth
to construct  the Voronoi tiling of 3D space, resulting in the convex polyhedron shown in panel (d).
Although the difference between the polyhedra in panels (c) and (d) is small, it remains clearly noticeable. Importantly, the same procedure of constructing a Voronoi tiling of space with respect to the nodes of the labyrinth, yields, for both Schwarz surfaces, the (e2) cell, which precisely matches the polyhedral approximation of the soft Voronoi cells.

Schoen \cite{Schoen} noticed this peculiar property for the gyroid, he
found the convex polyhedron in panel (d), identified it as the $L_2V_{17}$ polyhedron and called the corresponding
monohedral tiling a \emph{toy model} of the gyroid. Apparently, Schoen was aware of the difference between the polyhedra
in panels (c) and (d) of Figure \ref{fig:gyroid} and he also commented that the gyroid is not associated directly
with any convex tiling. 

In principle, the EEB algorithm can compute all soft cells that are first-order equivalents of a given polyhedral tiling. However, in the case of the nonconvex polyhedron shown in panel (c), this computation appears to be particularly challenging. Nonetheless, determining whether another soft cell exists that is first-order equivalent to the soft gyroid cell is an intriguing open question, given the significant role the gyroid structure plays in characterizing material microstructures \cite{grason}.

\subsection{Connecting the Schwarz surfaces}

As noted earlier, the Schwarz surfaces P and D represent the endpoints of a one-parameter family of minimal surfaces originally described by Bonnet \cite{Bonnet}. Such one-parameter families, in which one soft cell continuously transforms into another, are not merely mathematical curiosities; indeed, they have also been observed experimentally \cite{thomas}, highlighting their practical relevance in physical systems. The interpretation of both Schwarz surfaces as soft tilings (g2) and (i2) admits additional one-parameter connections between them, which we describe below.

Both (g2) and (i2) are first order equivalents of the (e2) tiling satisfying two additional constraints:  both result from the Voronoi decomposition of
tubular labyrinths and both are soft tilings. If we relax either the first or the second of the latter two properties then we  find  one-parameter families of first-order (e2) tilings that connect the Schwarz P to the Schwarz D surface.

First we relax the Voronoi construction: this is equivalent of relaxing the constraint of planar faces. This results in the connection (g2) $\to$ (h2) $\to$ (i2), consisting entirely of \emph{soft tilings}.
This connection, along the great circles $g_{abcd}$ and $g_{adbc}$ is marked by red lines in Figure \ref{fig:3}.
Next, we relax the softness condition and keep the constraint of planar faces.   This results in the connection (g2) $\to$ (e2) $\to$ (i2), consisting entirely of suitably defined \emph{Voronoi tilings}.
This connection, along the great circles $g_{quad}$ and $g_{hex2}$ is marked by blue lines in Figure \ref{fig:3}.
We may also relax \emph{both} conditions. In this case any one-parameter curve on the unit sphere passing through  (g2) and (i2) may be chosen. The shortest path is the great circle
 $g_{PD}$, marked by green line in Figure \ref{fig:3}, connecting the soft cells (g2) and (i2).  We computed  the unit normal as
\begin{equation}\label{Kelvin3}
\mathbf{u}_{PD}=\left(\frac{1}{\sqrt{12}},\frac{1}{\sqrt{12}},-\frac{2}{\sqrt{6}}
\right).
\end{equation}
While this solution appears to provide the most natural link between first-order (e2) tilings, it remains unclear  what specific geometric constraint it imposes on the shape of the corresponding cell. If we consider the intersection of the solutions $g_{PD}$ and $g_{hex1}$, we obtain an isolated solution, which we refer to as the \emph{PD-cell}. The coordinates of this cell are presented in Table \ref{tab:KelvinPD}. Notably, the PD-cell is not soft; the relevant scalar products can be explicitly calculated as $\mathbf{ab}=-3/7$, $\mathbf{ac}=1/7$, and $\mathbf{bc}=-5/7$.

\subsection{Soft tilings and symmetry groups}
Observation \ref{ob:schwarz} identified the soft cells generated by Voronoi partitions of TPMS and in 
Proposition \ref{prop:schwarz} we were able to identify the Voronoi cell of the Schwarz
P and Schwarz D surfaces with the soft cells (g2) and (i2),  both being first order equivalent to the truncated octahedron (e2) cell.
The soft tilings associated with these cells exhibit the full symmetry of their corresponding minimal surfaces, as their vertices align precisely with those of the regular map $\{6,4 |4\}$ described by Coxeter \cite{Coxeter}. Furthermore, TPMS accommodate various other regular maps and infinite polyhedra, each defining distinct soft monohedric tilings with reduced symmetry groups. These discrete structures thus enable the generation of soft monohedric tilings characterized by a range of smaller symmetry groups.

We also remark that, according to Theorem \ref{thm:0}, the soft tilings (f2) and (g2) inherit the full symmetry
group of the (e2) polyhedral tiling, the latter being one of the five
Dirichlet–Voronoi cells of point lattices \cite{ghorvath_dirichlet, voronoi_dirichlet}.
An examination of the remaining four Dirichlet-Voronoi cells shows that no corresponding soft tiling exists that preserves the full symmetry group of their polyhedral mosaics. An intriguing open question is whether, aside from the (f2) and (g2) tilings, there exist additional soft tilings that fully inherit the symmetry of their respective polyhedral tilings.

\subsection{Second order families of soft cells}
\begin{figure}[h!]
\begin{center}
\includegraphics[width=\columnwidth]{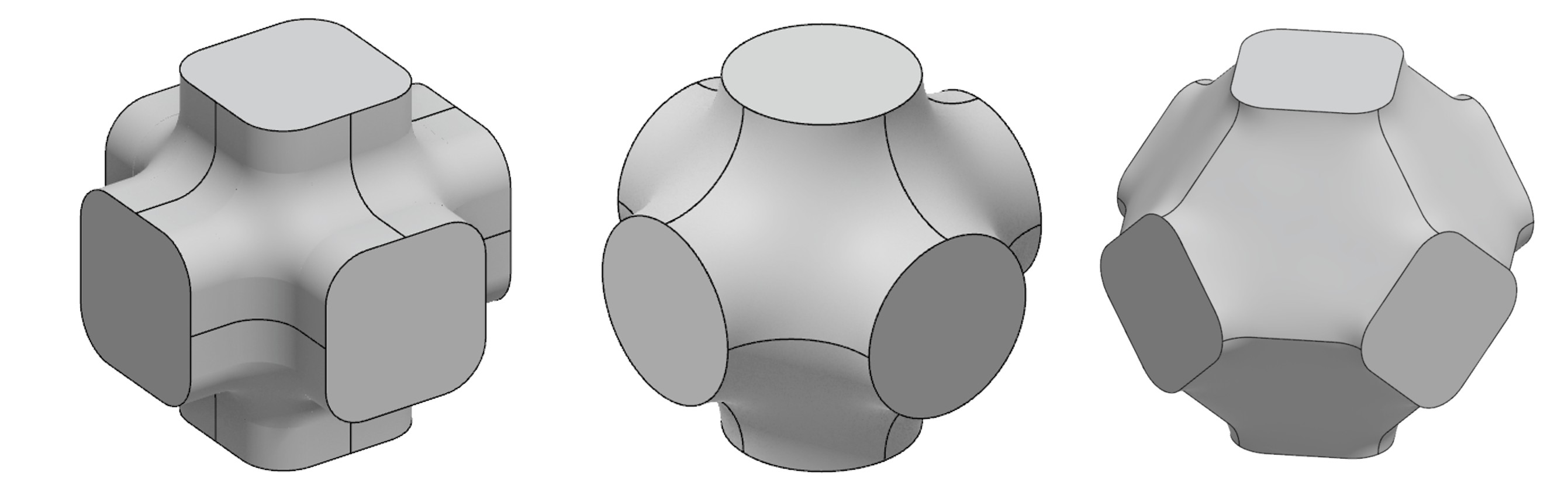}
\caption{Three members of the  second order family (g2) of soft cells, containing  also the Schwarz P unit cell}\label{fig:6}
\end{center}
\end{figure}

Although we only used its second order structure, the (e2) cell in Figure \ref{fig:1} is determined to fourth order
and the same holds for the Kelvin cell in Figure \ref{fig:5}.
In contrast,  the 4 soft cells (f2), (g2), (h2) and (i2) appearing in Theorem \ref{thm:1} and later in our computations
are, strictly speaking, only determined to second order. The images shown in Figures \ref{fig:1}, \ref{fig:3}
are examples of these second order families where we complemented the second order structure by circular edges and minimal surfaces as faces,
to obtain the full geometry of the cells. 
Indeed, the proof of Proposition \ref{prop:schwarz} relies precisely on the insight that specifying the exact geometric form of the Schwarz P and D cells is unnecessary to identify them as members of the (g2) and (i2) families. Visually, however, the differences between the Schwarz cells and the illustrated examples of the (g2) and (i2) families are minimal. To emphasize that our analysis is limited to second-order considerations and thus does not entirely constrain the cell shapes, we provide two additional examples of cells belonging to the (g2) family in Figure \ref{fig:6}.

\subsection{Summary}
In this paper, we introduced and described the EEB algorithm, which not only computes the shapes of soft cells up to the half-tangents of their edges, but also, in principle, is capable of identifying \emph{all} soft cells sharing the vertices of a given polyhedral tiling. To demonstrate the effectiveness of this algorithm, we applied it to the Dirichlet-Voronoi tiling of the $bcc$ lattice. We  proved that exactly two  types of soft cells exist, determined up to the half-tangents of their edges, which share the vertices and the full symmetry group of the polyhedral tiling. We also proved that exactly four types of soft cells exist, determined up to the half-tangents of their edges, that share the vertices of this monohedral tiling, exhibit at least tetrahedral symmetry, and possess at least one planar face. Furthermore, we established that two of these soft cells correspond precisely to the Voronoi cells associated with the Schwarz P and Schwarz D minimal surfaces, respectively.

The EEB algorithm has successfully identified soft cells that differ fundamentally from the standard examples previously illustrated in \cite{softcells1}. This discovery significantly expands the spectrum of potential applications and invites exciting new explorations into the diversity of soft-cell geometries yet to be uncovered.

\section*{Acknowledgement}
The authors sincerely thank Greg Grason for the stimulating, in-depth discussion on the geometry of mesoatoms.
KR and GD: This research was supported by NKFIH grants K149429 and EMMI FIKP grant VIZ. KR: This research has been supported by the program UNKP-24-3 funded by ITM and NKFI. The research was also supported by the Doctoral Excellence Fellowship Programme (DCEP) funded by ITM and NKFI and the Budapest University of Technology and Economics. The gift representing the Albrecht Science Fellowship is gratefully appreciated. 
\bibliographystyle{unsrt}
\bibliography{soft}

\end{document}